\documentclass[]{aastex631} 

\usepackage{textcomp}  
\usepackage{comment}
\usepackage{subfigure}
\usepackage{booktabs}

\newcommand\cc{black}

\accepted{\textcolor{\cc}{January 16, 2022}}
\submitjournal{ApJ}

\shorttitle{KamLAND IBD \nue - GRB time-coincident search}
\shortauthors{KamLAND collaboration}

\graphicspath{{./}{figures/}}

\begin{document}
\title{A search for correlated low-energy electron antineutrinos in KamLAND with gamma-ray bursts}

\correspondingauthor{S.~N.~Axani}
\email{saxani@mit.edu}

\newcommand{\tohoku}{\affiliation{Research Center for Neutrino Science, Tohoku University, Sendai 980-8578, Japan}}
\newcommand{\fris}{\affiliation{Frontier Research Institute for Interdisciplinary Sciences, Tohoku University, Sendai 980-8578, Japan}}
\newcommand{\gppu}{\affiliation{Graduate Program on Physics for the Universe, Tohoku University, Sendai 980-8578, Japan}}
\newcommand{\tohokuRigaku}{\affiliation{Department of Physics, Tohoku University, Sendai 980-8578, Japan}}
\newcommand{\ipmu}{\affiliation{Institute for the Physics and Mathematics  of the Universe, The University of Tokyo, Kashiwa 277-8568, Japan}}
\newcommand{\osakarcnp}{\affiliation{Graduate School of Science, Osaka University, Toyonaka, Osaka 560-0043, Japan}}   
\newcommand{\osaka}{\affiliation{Research Center for Nuclear Physics (RCNP), Osaka University, Ibaraki, Osaka 567-0047, Japan}}
\newcommand{\tokushima}{\affiliation{Graduate School of Advanced Technology and Science, Tokushima University, Tokushima 770-8506, Japan}}
\newcommand{\tokushimaGakusei}{\affiliation{Graduate School of Integrated Arts and Sciences, Tokushima University, Tokushima 770-8502, Japan}}
\newcommand{\kyoto}{\affiliation{Department of Physics, Kyoto University, Kyoto 606-8502, Japan}}
\newcommand{\lbl}{\affiliation{Nuclear Science Division, Lawrence Berkeley National Laboratory, Berkeley, CA 94720, USA}}
\newcommand{\hawaii}{\affiliation{Department of Physics and Astronomy, University of Hawaii at Manoa, Honolulu, HI 96822, USA}}
\newcommand{\mituniv}{\affiliation{Massachusetts Institute of Technology, Cambridge, MA 02139, USA}}
\newcommand{\bu}{\affiliation{Boston University, Boston, MA 02215, USA}}
\newcommand{\tennessee}{\affiliation{Department of Physics and Astronomy,  University of Tennessee, Knoxville, TN 37996, USA}}
\newcommand{\tunl}{\affiliation{Triangle Universities Nuclear Laboratory, Durham, NC 27708, USA }}    
\newcommand{\chapehill}{\affiliation{The University of North Carolina at Chapel Hill, Chapel Hill, NC 27599, USA}}

\newcommand{\northcarolina}{\affiliation{North Carolina Central University, Durham, NC 27701, USA}}
\newcommand{\duke}{\affiliation{Department of Physics, Duke University, Durham, NC 27705, USA}}
\newcommand{\seattle}{\affiliation{Center for Experimental Nuclear Physics and Astrophysics, University of Washington, Seattle, WA 98195, USA}}
\newcommand{\nikhef}{\affiliation{Nikhef and the University of Amsterdam, Science Park,  Amsterdam, the Netherlands}}
\newcommand{\virginia}{\affiliation{Center for Neutrino Physics, Virginia Polytechnic Institute and State University, Blacksburg, VA 24061, USA}}



\author{S.~Abe}\tohoku
\author{S.~Asami}\tohoku
\author{A.~Gando}\tohoku
\author{Y.~Gando}\tohoku
\author{T.~Gima}\tohoku 
\author{A.~Goto} \tohoku
\author[0000-0002-4238-7990]{T.~Hachiya}\tohoku
\author{K.~Hata} \tohoku
\author{K.~Hosokawa} \tohoku
\author[0000-0001-9783-5781]{K.~Ichimura} \tohoku  
\author[0000-0001-7694-1921]{S.~Ieki} \tohoku
\author{H.~Ikeda}\tohoku
\author{K.~Inoue}\tohoku \ipmu 
\author[0000-0001-9271-2301]{K.~Ishidoshiro}\tohoku
\author{Y.~Kamei} \tohoku
\author[0000-0003-2350-2786]{N.~Kawada} \tohoku
\author{Y.~Kishimoto} \tohoku \ipmu
\author{T.~Kinoshita} \tohoku 
\author{M.~Koga}\tohoku \ipmu 
\author{N.~Maemura}\tohoku
\author{T.~Mitsui}\tohoku
\author{H.~Miyake}\tohoku
\author{K.~Nakamura}\tohoku 
\author{K.~Nakamura}\tohoku 
\author{R.~Nakamura}\tohoku
\author{H.~Ozaki}\tohoku \gppu
\author{T.~Sakai} \tohoku 
\author{H.~Sambonsugi}\tohoku
\author{I.~Shimizu}\tohoku
\author[0000-0002-3988-2309]{J.~Shirai}\tohoku
\author{K.~Shiraishi}\tohoku
\author{A.~Suzuki}\tohoku
\author{Y.~Suzuki}\tohoku 
\author{A.~Takeuchi}\tohoku
\author{K.~Tamae}\tohoku

\author{M.~Eizuka}\tohoku
\author{M.~Kurasawa}\tohoku
\author{T.~Nakahata}\tohoku
\author{S.~Futagi}\tohoku

\author[0000-0002-2363-5637]{H.~Watanabe}\tohoku
\author{Y.~Yoshida} \tohoku
\author[0000-0003-3488-3553]{S.~Obara}\fris 
\author{A.~K.~Ichikawa} \tohokuRigaku
\author{S.~Yoshida}\osakarcnp
\author{S.~Umehara}\osaka
\author{K.~Fushimi}\tokushima

\author{B.~E.~Berger}\lbl \ipmu
\author[0000-0002-7001-717X]{B.~K.~Fujikawa}\lbl \ipmu
\author{J.~G.~Learned}\hawaii
\author{J.~Maricic}\hawaii
\author{S.~N.~Axani}\mituniv
\author{J.~Smolsky}\mituniv
\author{C.~Laber-Smith}\mituniv
\author{L.~A.~Winslow}\mituniv
\author{Z.~Fu}\mituniv
\author{J.~Ouellet}\mituniv 
\author{Y.~Efremenko}\tennessee \ipmu
\author{H.~J.~Karwowski}\tunl \chapehill
\author{D.~M.~Markoff}\tunl \northcarolina
\author{W.~Tornow}\tunl \duke \ipmu
\author[0000-0002-4844-9339]{A.~Li}\chapehill
\author{J.~A.~Detwiler}\seattle \ipmu
\author{S.~Enomoto}\seattle \ipmu
\author[0000-0002-1577-6229]{M.~P.~Decowski}\nikhef \ipmu
\author{C.~Grant}\bu
\author{H.~Song}\bu
\author{T.~O'Donnell}\virginia
\author{S.~Dell'Oro}\virginia

\newcommand{\nue}{$\bar{\nu}_e$ }
\newcommand{\ANALYSISSTARTDATE}{Sept. 3rd, 2002 }
\newcommand{\ANALYSISENDDATE}{xxx. xxxrd, 2021 }

\collaboration{99}{(KamLAND Collaboration)}

\begin{abstract}
We present the results of a time-coincident event search for low-energy electron antineutrinos in the KamLAND detector with gamma-ray bursts from the Gamma-ray Coordinates Network and Fermi Gamma-ray Burst Monitor. Using a variable coincidence time window of $\pm$500\,s plus the duration of each gamma-ray burst, no statistically significant excess above background is observed. We place the world's most stringent 90\% confidence level upper limit on the electron antineutrino fluence below \textcolor{\cc}{17.5\,MeV}. Assuming a Fermi-Dirac neutrino energy spectrum from the gamma-ray burst source, we use the available redshift data to constrain the electron antineutrino luminosity and effective temperature.
\end{abstract}



\section{Introduction} \label{sec::intro}


While gamma-ray bursts (GRB) represent some of the most luminous electromagnetic sources of radiation in the known universe, their progenitors have long been a mystery. 
This has been partially resolved with the recent observation of a gravitational-wave signal (GW170817) correlated with the nearest observed GRB to date (GRB170817A)~\citep{abbott2017gw170817,ligo2017multi}, appearing to confirm the hypothesis that the class of ``short'' GRBs (SGRB) originate from binary mergers~\citep{nakar2007short}. The class of ``long'' GRBs (LGRB), by comparison, are found in host galaxies with more active star formation and are thought to result from the core-collapse of massive stars, 
exemplified by the correlated observation of GRB130427A with SN2013cq~\citep{xu2013discovery}
. In either case, the progenitors are expected to release copious amounts of energy in the form of neutrinos ($\mathcal{O}$($10^{53}$\,erg)). 
The neutrino emission from the toroidal 
accretion disk surrounding the remnant of the binary star merger is expected to be the dominant source of MeV-scale neutrinos~\citep{rosswog2003high,
setiawan2006three}, with subdominant flux contribution from the hot dense remnant itself~\citep{ruffert1996coalescing}. This mechanism is similar to the models which also produce neutrinos in Type-II supernovae, although the neutron-rich environment of the binary star merger is expected to produce a larger electron antineutrinos ($\bar{\nu}_e$) flux~\citep{rosswog2003high}. 

The Kamioka Liquid scintillator Anti Neutrino Detector (KamLAND) has previously searched for astrophysical neutrinos associated with gravitational waves\,\citep{gando2016searchGW,abe2021search}, solar flares\,\citep{abe2021searchsolar}, and supernova~\citep{abe2021limits}. This paper represents an updated search to the KamLAND result presented in \cite{asakura2015study}. Time-coincident GRBs with MeV-scale neutrinos searches have also been performed by the Super-Kamiokande~\citep{fukuda2002search}, Borexino~\citep{agostini2017borexino}, and SNO~\citep{aharmim2014search}. 
High-energy neutrinos are also expected to accompany GRBs in the collimated jet structure that forms from accreting matter. These searches have been performed by the IceCube~\citep{abbasi2009search,abbasi2010search},
ANTARES~\citep{albert2017search}, AMANDA~\citep{hughey2005neutrino}, and Baikal~\citep{avrorin2011search} collaborations.

In this paper, we present the results of a search for $\bar{\nu}_e$s with energies ranging from $E_{\bar{\nu}_e} = 1.8$\,MeV to 100\,MeV contained in the \textcolor{\cc}{4931.1}\,days livetime (\textcolor{\cc}{8.6}\,kiloton-year) dataset from KamLAND, coincident with \textcolor{\cc}{2795} LGRBs and \textcolor{\cc}{465} SGRBs. We then use the subset of GRBs which have a measured redshift to constrain the \nue source luminosity and effective temperature.  


\section{KamLAND detector} \label{sec:detector} 

\begin{figure}[t]
    \centering
    \gridline{
        \fig{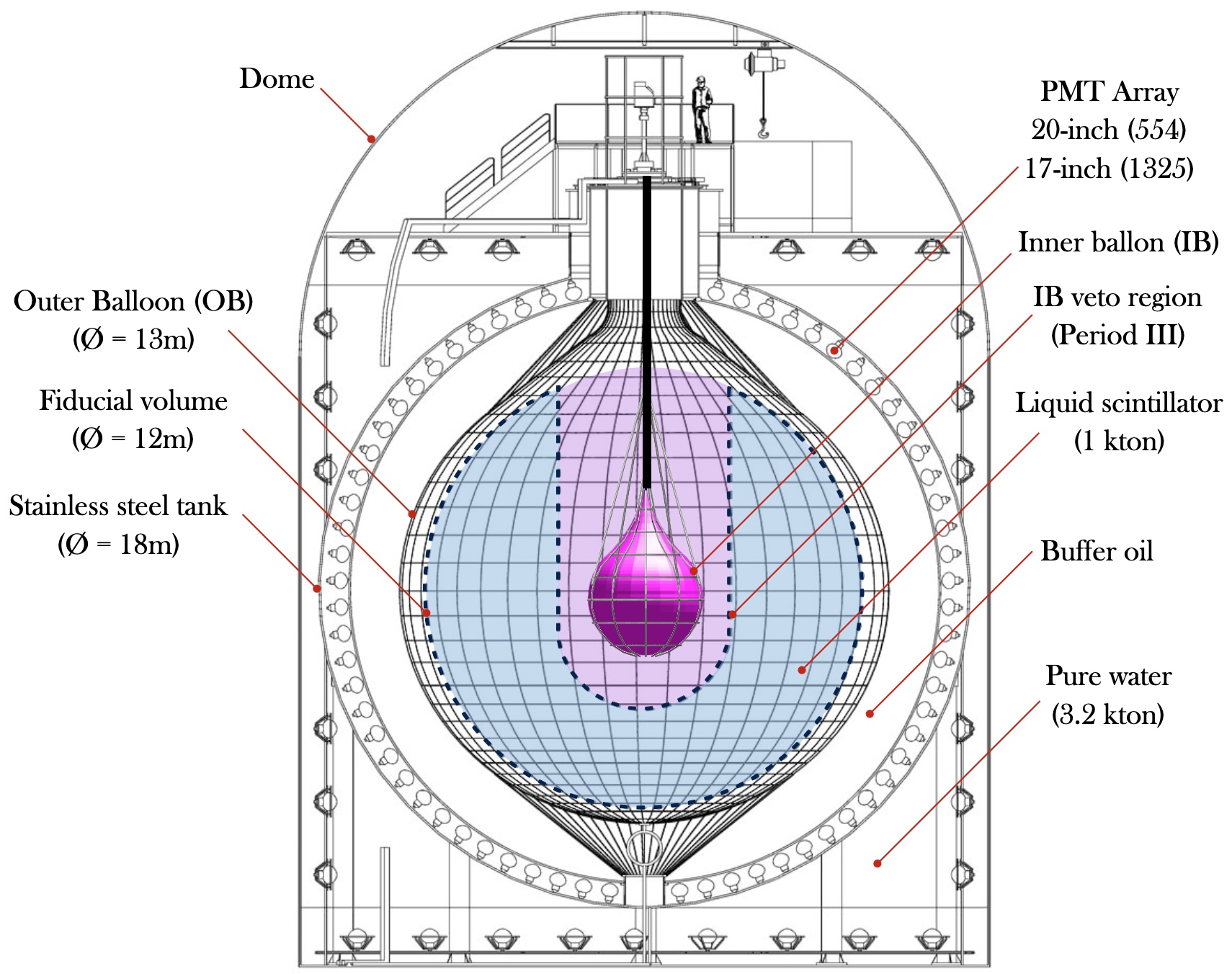}{.55\textwidth}{}
        \fig{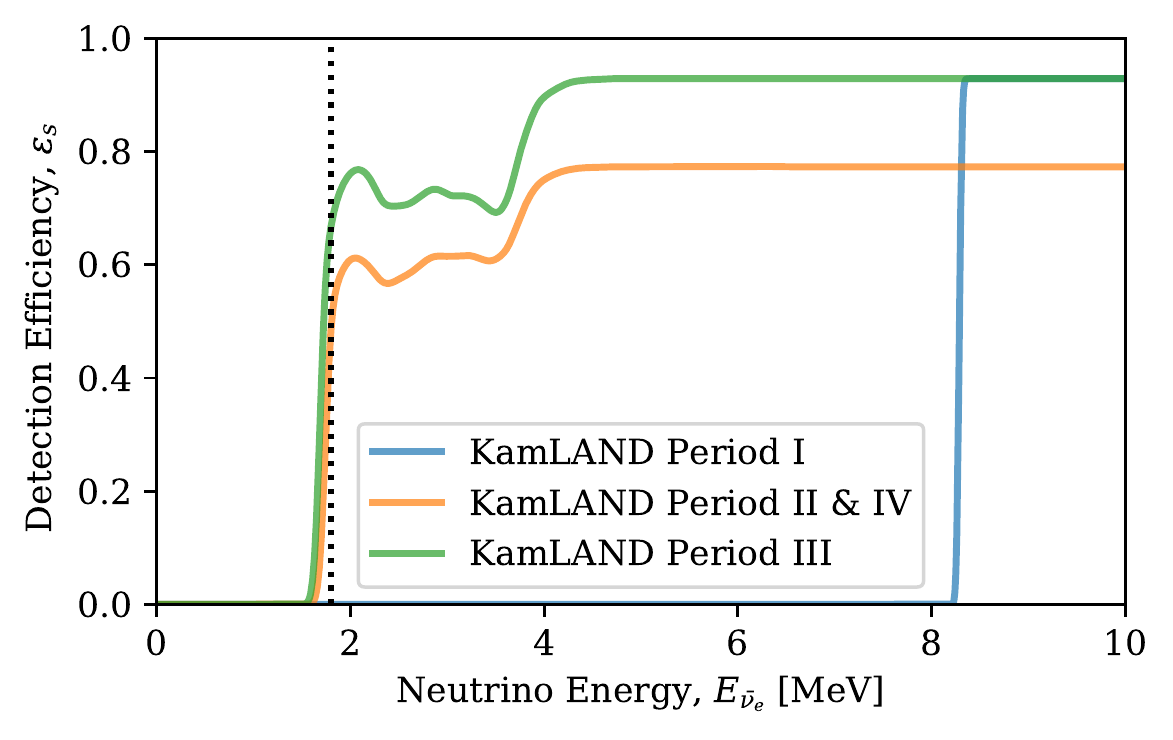}{0.44\textwidth}{}
    }
    \caption{Left: A schematic diagram of the KamLAND detector. The fiducial volume is highlighted in blue. We illustrate the azimuthally-symmetric inner balloon cut for data Period II \& IV in purple.  Right: The IBD \nue selection efficiencies for the four KamLAND periods. 
    The structure below $\sim$4\,MeV arises from the likelihood selection. A vertical dotted line is shown at the 1.8\,MeV low-energy IBD threshold. The non-zero efficiency below 1.8\,MeV is a result from smearing due to the energy resolution of the KamLAND detector.}
\label{fig::detector}
\end{figure}

A schematic diagram of the KamLAND neutrino detector is shown in Fig.~\ref{fig::detector} (left). The detector itself is situated in Kamioka, Japan, approximately 1\,km under the surface of Mt.\,Ikenoyama.  The detector is divided into two major sections: the inner detector and the outer detector, separated by an 18.0\,m diameter spherical stainless steel tank. The inner detector was optimized for low-energy \nue interactions, primarily to measure geo-neutrinos and neutrino oscillations using reactor neutrinos~\citep{gando2013reactor,araki2005experimental,abe2008precision}. 
 
Inside the inner detector, there is a 13.0\,m diameter spherical balloon made of a five layer EVOH and nylon film, 135\,$\mu$m thick.  This balloon holds 1000\,tons (1171\,$\pm$\,25\,m$^3$) of ultra pure liquid scintillator, which is by volume 80\% dodecane ($\rho = 0.75\,{\mathrm{g}\, \mathrm{cm}^{-3}}$) and 20\% pseudocumene ($\rho = 0.88\,{\mathrm{g}\, \mathrm{cm}^{-3}}$). 
The inner 12.0\,m diameter region is used as the fiducial volume for \nue detection. Outside of the balloon, the inner detector is filled with a buffer oil which helps to filter out gamma radiation from radioactive impurities in the detector or surrounding material. The outer shell of the inner detector holds an array of 1325 17-inch photomultipliers (PMT) and 554 20-inch PMTs pointed radially inwards.  These PMTs provide a total detector photocathode coverage of 34\%. The vertex resolution of reconstruction is $\sim$12\,cm/$\sqrt{E (\text{MeV})}$, and the energy resolution is $\sim$6.4\%/$\sqrt{E\,(\text{MeV})}$~\citep{gando2013reactor}.

The readout electronics are synchronized with the Global Positioning System (GPS) via a pulse-per-second trigger from a receiver placed outside the entrance to the Kamioka mine. Based on the uncertainty from the accuracy from the GPS system, signal transportation into the mine, optical/electrical signal conversion, and trigger system uncertainties, we estimate the total uncertainty in the absolute event timestamp in KamLAND to be $\mathcal{O}$(100\,$\mu$s). 

From 2011 October to 2015 October, a tear-drop shaped, 3.08\,m diameter, 25\,$\mu$m thick, nylon inner balloon was installed in the center of the detector. During this period, the inner balloon was filled with approximately 326\,kg of $^{136}$Xe loaded liquid-scintillator for the KamLAND-Zen\,400 neutrinoless double-beta decay experiment~\citep{gando2016search}. From 2018 April onward, a larger (3.8\,m diameter) inner balloon was inserted into the detector and filled with approximately 745\,kg of enriched Xenon (676\,kg of $^{136}$Xe) for the KamLAND-Zen\,800 experiment~\citep{gando2020first, gando2021nylon}.

\section{Data selection and background estimation} \label{sec:antiv}

\subsection{Electron antineutrino selection}
We use the inverse beta-decay (IBD) reaction ($\bar{\nu}_e + p \rightarrow e^+ + n$) to search for \nue interactions in the KamLAND detector. The selection criteria for these events is described in detail in \citet{gando2013reactor,Abe_2021}. The IBD reaction has a threshold of $E_{\bar{\nu}_e} = 1.803$\,MeV and creates a delayed-coincidence (DC) event pair, which can be identified by both the separation in time (0.5\,$\mu\mathrm{s}\leq \Delta T \leq 1000$\,$\mu$s) and space separation ($\Delta R$\,$\leq 200$\,cm) between the prompt $e^+$ thermalization then annihilation and delayed neutron capture. The mean capture time for the neutron is $207.5\pm2.8$\,$\mu$s. The DC identification method greatly suppresses backgrounds and allows for a high \nue detection efficiency. The prompt energy, $E_p$, is equal to the sum of the $e^+$ kinetic energy and its annihilation energies ($E_p = T_{e_+} + 2m_e$). We can relate the prompt energy to the incident neutrino through: $E_{\bar{\nu}_e} = E_p + \delta E + T_n$. Here, $T_n$ is the kinetic energy of the neutron  and $\delta E$ is equal to 0.782\,MeV.  The thermalization of the neutron is also contained in the prompt event, however it is quenched and can be ignored for low energies  ($E_{\bar{\nu}_e} < 20$\,MeV), therefore throughout this analysis, we will use:
\begin{equation}
\label{eq::Ep_conversion}
E_{\bar{\nu}_e} \approx E_p + \delta E.
\end{equation}
This approximation is sufficiently accurate for the results presented in this paper, and represents an 10\% bias at $E_{\bar{\nu}_e} = 100$\,MeV~\citep{asakura2015study}.

The delayed neutron in the DC pair interacts with a $^1$H\,($^{12}$C) molecule to form a deuteron\,($^{13}$C) and 2.22\,MeV\,(4.95\,MeV) gamma-ray. We include a cut on the reconstructed delayed energy, $E_d$, such that $1.8 \mathrm{MeV} \leq E_d \leq 2.6$\,MeV or 4.4\,MeV\,$\leq E_d \leq 5.6$\,MeV. 

An additional likelihood-based selection is used to distinguish \nue DC pairs from coincident background events. The likelihood selection accounts for the accidental coincidence rates, the outer detector refurbishment~\citep{ozaki2017refurbishment}, the inner balloon installation~\citep{GandoY_2020, gando2021nylon}, and the status of some of the reactors in Japan. Details of this likelihood selection method are further described in 
\citet{gando2013reactor}.

We include KamLAND IBD data spanning from the first GRB in the catalog, 2004\,December\,19th, to the most recently verified KamLAND data taking run, 2021\,June\,12th, separated into four time periods. Period I: This period includes all KamLAND data prior to the installation of the KamLAND-Zen\,400 inner balloon in 2011 October. The \nue selection includes a low-energy threshold of $E_p$\,=\,7.5\,MeV to reduce the Japanese nuclear reactor neutrinos background. Following a major earthquake in 2011 March, the operation of all nuclear power plants in Japan was suspended and therefore all subsequent time periods in this analysis have a low energy threshold of $E_p$\,=\,0.9\,MeV. Period II: This spans the time in which the inner balloon was inserted into the KamLAND detector for the KamLAND-Zen\,400 experiment (2011 October to 2015 October). Here, we include an additional geometric cut on the delayed event around the inner balloon to reduce backgrounds from the inner balloon material. The cut removes the central spherical 2.5\,m radius region, extending to the top of the detector in a 2.5\,m radius cylinder. Period III: This period begins at the time of removal of the KamLAND-Zen\,400 inner balloon, and spans the time up until the introduction of the KamLAND-Zen\,800 inner balloon. Period IV: This is during the KamLAND-Zen\,800 experiment and reintroduces the same geometric cut described in Period II. 

The inner balloon geometric cuts for Period II \& IV are illustrated as the purple region in Fig.~\ref{fig::detector} (left), while the IBD selection efficiency as a function of \nue energy for each time period is shown in Fig.~\ref{fig::detector} (right). Above 10\,MeV, the selection efficiencies converge to 92.9\% and 77.4\% for Periods I\,\&\,III and II\,\&\,IV, respectively. The reduction in fiducial volume due to the geometric cuts for Period II \& IV are accounted for in the selection efficiencies rather than in the number of target nuclei.


\subsection{GRB event selection}\label{sec::GRB}

The GRB events are extracted from GRBWeb~\citep{aguilar2011online}, an online cataloging tool which parses the Gamma-ray burst Coordinate Network\footnote{\url{https://gcn.gsfc.nasa.gov}} (GCN) publicly available circulars and the Fermi Gamma-ray Burst Monitor\footnote{\url{https://gammaray.nsstc.nasa.gov/gbm/}} (GBM). The circulars archive reports from satellites such as SWIFT, Fermi, INTEGRAL, HETE-2, AGILE, Ulysses, Suzaku, and WIND/Konus, along with supplemental data from ground-based observatories. We require that each event has an absolute trigger time ($t_{GRB}$) and a measured GRB duration ($t_{90}$), which represents the time interval in which the integrated photon counts increase from 5\% to 95\% of the total counts. We separate the GRBs into two classes demarcated by their duration: GRBs with $t_{90} < 2$ seconds are labelled ``short,'' whereas those with $t_{90} \geq 2$ are ``long''~\citep{kouveliotou1993identification}. After imposing a quality check on the KamLAND data, ensuring that we do not include GRBs that arrive during KamLAND detector deadtime (e.g. during calibration runs or non-stable operation), the final GRB event selection sample contains \textcolor{\cc}{2795} LGRBs and \textcolor{\cc}{465} SGRBs. Of these, \textcolor{\cc}{377} LGRBs and \textcolor{\cc}{34} SGRBs are found to have a measured redshift. We will refer to the sum of the LGRB and SGRB datasets as the ``combined set.'' Of note, contrary to classification scheme presented here, although GRB170817A (mentioned in Sec.~\ref{sec::intro}) had a measured duration of $t_{90} = 2.048$\,s we have placed it in the SGRB dataset.
A full list of the GRBs used in this analysis can be found on our website\footnote{\url{https://www.awa.tohoku.ac.jp/kamland/GRB/2021/index.html}}.

\subsection{Background estimation}
The background rate is calculated independently for each of the four KamLAND periods. The IBD events which occur outside of the coincidence time window around each GRB are used to determine the uncorrelated background rate. In Period I, enforcing that $E_p > 7.5$\,MeV, the background is dominated by long-lived spallation products and fast neutrons from cosmic-ray muons~\citep{abe2010production}, and neutral current (NC) atmospheric neutrino interactions. While this is also the dominant source of higher energy neutrinos in Period II\,-\,IV, the dominant neutrino sources below approximately 8\,MeV are the Japanese nuclear reactor power plants and geo-neutrinos from radioactive decays in the Earth (primarily below 3.4\,MeV). Other backgrounds include DC pairs induced by the decay of radioactive impurities, spallation-produced $^{9}$Li and $^8$He leading to a $\beta$-decay followed by a neutron capture~\citep{abe2010production}, and alpha-induced $^{13}$C($\alpha$,n)$^{16}$O reaction in the liquid scintillator. Using the background rates in each period and the sum of all coincidence time windows, the expected number of background events is found to be \textcolor{\cc}{2.94} for the LGRB and \textcolor{\cc}{0.47} for the SGRB. The period dependent background information is shown in Table~\ref{table::PeriodData}.



\begin{deluxetable*}{ccccccccccc}[h]
    \color{\cc}
    \label{table::PeriodData}
    \tablecaption{The period dependent information used in this analysis. 
    The livetime efficiency, $\epsilon_{\mathrm{live}}$, is defined as the ratio between the livetime and realtime. The total number of neutrinos observed per period, along with information regarding the number of observed SGRBs and LGRBs, their total time windows, 
    and the expected number of background events are also shown.}
    \tablewidth{0pt}
    \tablehead{
        {} &  \colhead{Realtime} & \colhead{Livetime} & \colhead{$\epsilon_{\mathrm{live}}$} & \colhead{\nue} & \multicolumn3c{Short Gamma-ray Bursts} & \multicolumn3c{Long Gamma-ray Bursts} \\ 
        \cmidrule(lr){6-8}\cmidrule(lr){9-11}
        {} & \colhead{[days]} & \colhead{[days]} & \colhead{[\%]} & \colhead{Counts} & \colhead{Counts} & \colhead{Window [h]} & \colhead{Bkg. Exp.} & \colhead{Counts} & \colhead{Window [h]} & \colhead{Bkg. Exp.}
    }
    \startdata
    Period I   & 2487.4 & 1985.0 & 79.8 & 14 & 121 & 33.7 & 0.01  & 838 & 262.1  & 0.08 \\ [0.5ex]
    Period II  & 1474.0 & 1371.8 & 93.1 & 148 & 172 & 47.8 & 0.22  & 911 & 279.5  & 1.26 \\ [0.5ex]
    Period III & 558.9 & 480.3 & 85.9 & 58 & 56 & 15.6 & 0.08  & 323 & 97.7  & 0.49 \\ [0.5ex]
    Period IV  & 1152.7 & 1094.0 & 94.9 & 132 & 116 & 32.3 & 0.16  & 723 & 221.9  & 1.12 \\ [0.5ex]
    \hline
    Total      & 5673.0 & 4931.1 & 86.9 & 352 & 465 & 129.3 & 0.47  & 2795 & 861.2  & 2.94 \\ [0.5ex]
    \hline
    \enddata
 \color{black}
\end{deluxetable*}

\section{Time-Coincident Event Search} \label{sec:candidates}
We perform a time-coincident analysis searching for IBD events that coincide with a predefined window size around each GRB. The window size for each GRB trigger time, $t_{GRB}$, is defined such that:
\begin{equation}
t_{GRB} - t_{p} < t_{DC} < t_{GRB} +t_{p} +t_{90},
\label{eq::window}
\end{equation} 
where $t_{DC}$ is the DC IBD event time and the variable $t_{p}$ represents a  predefined window size of $\pm$500\,s. $t_{p}$ is chosen to be sufficiently large to cover reasonable model dependent time differences between the neutrino and photon production within the GRB, and the neutrino time-of-flight delay. The model dependent time differences account for temporal effects from the source evolution, such as those originating from precursor activity~\citep{lazzati2005precursor,burns2020neutron}, accretion disk evaporation lifetimes~\citep{beloborodov2008hyper}, and core-collapse and Kelvin Helmholtz cooling phase time-scales~\citep{totani1998future}.

Due to the non-zero neutrino mass, a neutrino time-of-flight delay is expected relative to the photon travel time. Using the neutrino oscillation parameters from~\citet{esteban2020fate} and assuming the sum of the neutrino masses to be $<1.2$\,eV, we calculate the most massive neutrino state to be $<59$\,meV, and conservatively set this to be the mass of \nue. Then, using the base-$\Lambda$CDM cosmological parameters from \citet{aghanim2020planck}, the time-of-flight delay for a 1.8\,MeV \nue from the most distant GRB (GRB100205A, with a suspected redshift z\,$<13$ \citep{chrimes2019case, kim2012near}) is estimated to be approximately 129\,s. Therefore the $t_{p} = \pm500$\,s is considered conservative.

The sum of all time windows, from Eq.~(\ref{eq::window}), is found to be \textcolor{\cc}{861.2}\,h for LGRBs, \textcolor{\cc}{129.3}\,h for SGRBs. The breakdown between periods is shown in Table~\ref{table::PeriodData}.





\section{Results}
We find a single coincident event between a long gamma-ray burst, GRB180413A, and a low-energy IBD interaction in KamLAND during Period III. This GRB was extracted from the Fermi GBM and found to have a duration of $t_{90} = 57.86\pm2.36$\,s~\citep{aguilar2011online}. This LGRB does not have a measured redshift. The coincident \nue was found to have a prompt (delayed) energy  of $E_{p} = 4.24 \pm0.13$\,MeV ($E_{d} = 2.28 \pm0.10$\,MeV) and arrived 191.3\,s prior to GRB180413A. This observation is consistent with the background expectation.


The 90\% confidence level (C.L.) lower and upper limits on the number of GRB-correlated IBD events is calculated according to the Feldman-Cousins procedure~\citep{feldman1998unified}. For the LGRB, SGRB, and combined set, we find the 90\% C.L. intervals to be \textcolor{\cc}{[0.0, 1.91], [0.0, 1.97], and [0.0, 1.64]} signal events, respectively. The upper limits of each interval, $N_{90}$, can be used to place an upper limit on the \nue fluence per GRB:

\begin{figure}[t]
\centering
\includegraphics[width=0.7\columnwidth]{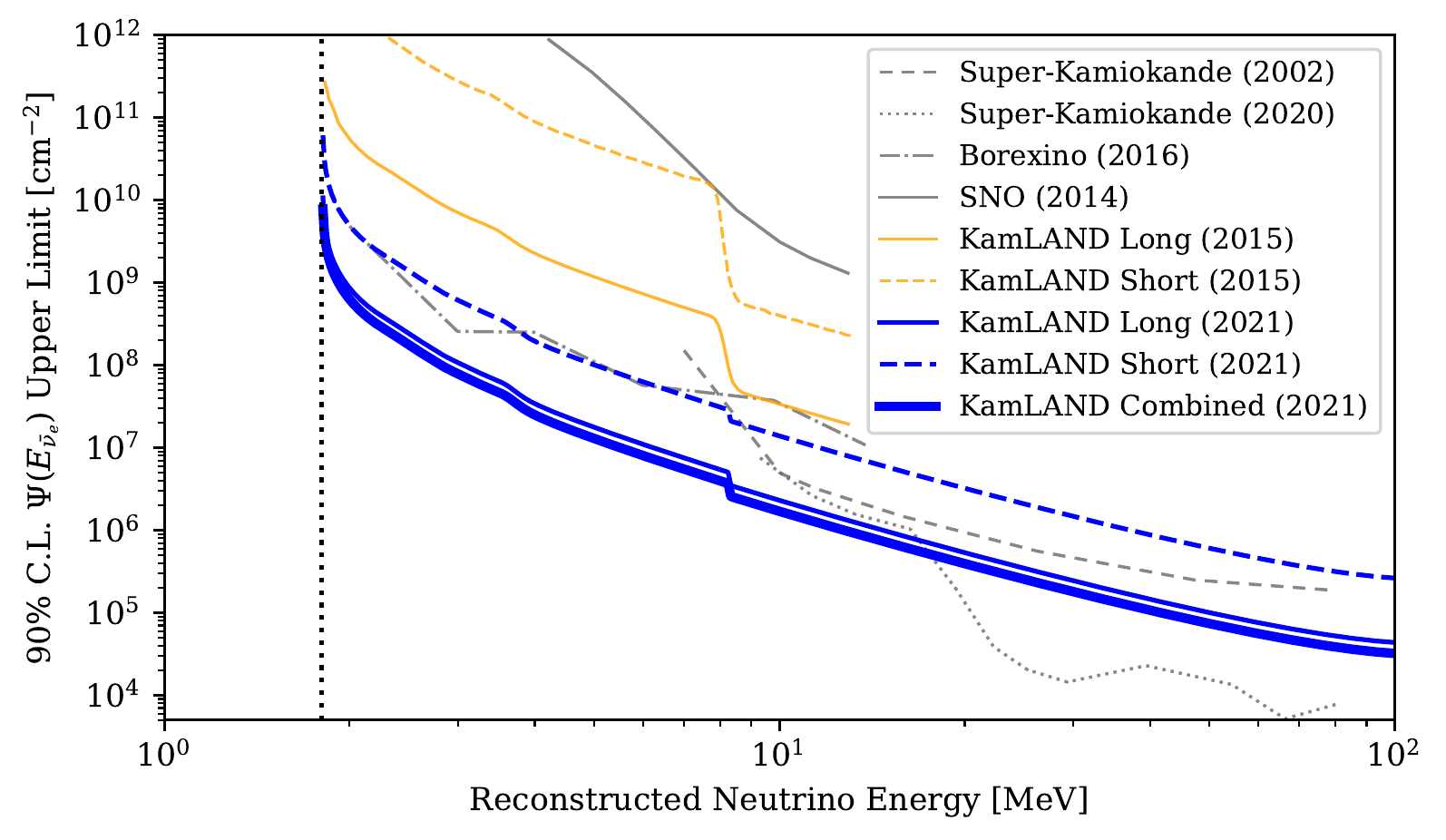}
\caption{
The 90\% C.L. \nue fluence limit Green’s function as a function of neutrino energy. The GRB combined set, shown in thick solid blue is compared to results from Super-Kamiokande~\citep{fukuda2002search,orii2021search} Borexino~\citep{agostini2017borexino}, and SNO~\citep{aharmim2014search}. Below 17.5\,MeV, KamLAND establishes the tightest limits on \nue fluence. The slight distortion around $E_{\bar{\nu}_e} \approx 3$\,MeV results from the energy dependence of the likelihood selection efficiency. Similarly the bump at $E_{\bar{\nu}_e} \approx  8.3$\,MeV comes from the prompt energy cut introduced in Period I. This figure also shows the LGRB and SGRB upper limits in thin solid and dashed blue, to be compared to the previous KamLAND result shown in orange. A vertical dotted line is shown at 1.8\,MeV, the low-energy IBD threshold.
}
\label{fig::fluence}
\end{figure}

\begin{equation}\label{eq::fluence}
F_{90} = \frac{N_{90}} {N_T  \sum_{k=1}^{k=4} \epsilon_{live}^k  N^k_{GRB} \int  \sigma(E_{\bar{\nu}_e}) \lambda(E_{\bar{\nu}_e})  \epsilon^k_s(E_{\bar{\nu}_e}) dE_{\bar{\nu}_e} },
\end{equation}
where $N_{T}$ = 5.98$\times10^{31}$ is the number of target nuclei in fiducial volume, $\sigma$($E_{\bar{\nu}_e}$) is the IBD cross-section~\citep{strumia2003precise}, and $\lambda (E_{\bar{\nu}_e})$ is the normalized \nue energy spectrum. The summation iterates over the periods $k$, where $\epsilon_{live}^k$ represent the livetime efficiencies, $N_{GRB}^k$ are the number of observed GRBs, and $\epsilon^k_s$ are the selection efficiencies from Fig.~\ref{fig::detector} (right). The \nue energy spectrum, $\lambda (E_{\bar{\nu}_e})$, is often modelled as a temperature dependent Fermi-Dirac (FD) distribution with zero chemical potential for core-collapse supernovae and binary mergers~\citep{horiuchi2009diffuse,gando2016searchGW}:
\begin{equation}
\lambda(E_{\bar{\nu}_e},T)_{FD} =  \frac{1}{T^3 f_2}\frac{E_{\bar{\nu}_e}^2}{e^{(E_{\bar{\nu}_e}/T)}+1}, \quad f_n = \int_0^{\infty} \frac{x^n}{e^{x} +1} dx.
\end{equation}
If we take the mean \nue energy to be \textlangle{}$E$\textrangle{}\,$= 12.7$\,MeV and set $T\,=\,$\textlangle{}$E$\textrangle{}$/3.15$, we find the 90\% C.L. upper limit on the \nue fluence, integrating Eq.~(\ref{eq::fluence}) from 1.8\,MeV to 100\,MeV, per GRB to be:

\begin{equation}\label{f90}
F_{90}^{\mathrm{LGRB}} = 1.03\times 10^6 \,\mathrm{cm}^{-2} ,\quad
F_{90}^{\mathrm{SGRB}} = 6.27\times 10^7 \,\mathrm{cm}^{-2} ,\quad
F_{90}^{\mathrm{Combined}} = 0.75\times 10^6 \,\mathrm{cm}^{-2}.
\end{equation}
Alternatively, without making any assumption on the \nue energy spectrum, we instead calculate the equivalent model independent Greene's Function, $\Psi(E_{\bar{\nu}_e})$, comparable to \citet{fukuda2002search}, by setting $\lambda (E_{\bar{\nu}_e}) = \delta (E_{\bar{\nu}_e} - E_{\bar{\nu}_e}^{'})$:

\begin{equation}\label{eq::greenes}
\Psi(E_{\bar{\nu}_e}) = \frac{N_{90}} {N_T \sigma(E_{\bar{\nu}_e})  \Sigma_{k=1}^{k=4} \epsilon_{live}^k N^k_{GRB}   \epsilon^k_s(E_{\bar{\nu}_e}) }.
\end{equation}

The results of Eq.~(\ref{eq::greenes}) for the LGRBs (thin solid blue), SGRBs (thin dashed blue), and combined set (thick blue) are presented in Fig.~\ref{fig::fluence}.
We find the combined upper limit reaches a minimum of approximately \textcolor{\cc}{3.2}$\times 10^4$\,cm$^{-2}$ at $E_{\bar{\nu}_e}$\,=\,100\,MeV. This result represents the world's most stringent limits for \nue energies below \textcolor{\cc}{17.5\,MeV}. The sensitivities of this measurement are found to be \textcolor{\cc}{2.3, 1.4, 2.7} times larger than the observed limits for the LGRBs, SGRBs, and combined sets, respectively.  The improvement over the previous KamLAND result~\citep{asakura2015study}, represented by solid and dashed orange lines, originates from the increase in GRB statistics and the under-fluctuation in the background. 

\vspace*{\fill}
\subsection{Source \nue luminosity-temperature constraints}\label{sec::lum}
\begin{figure}[t]
    \centering
    \gridline{
        \fig{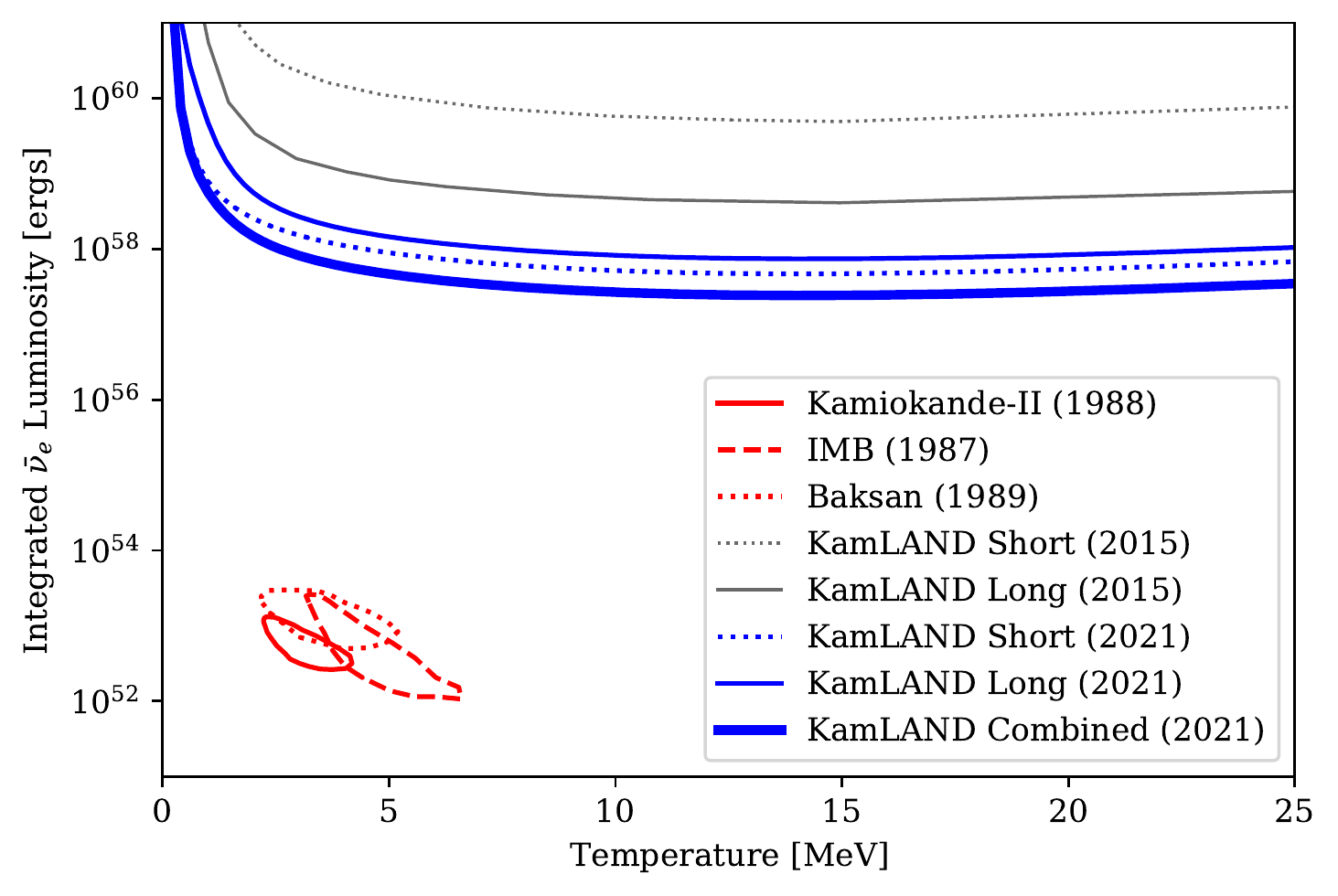}{.486\textwidth}{}
        \fig{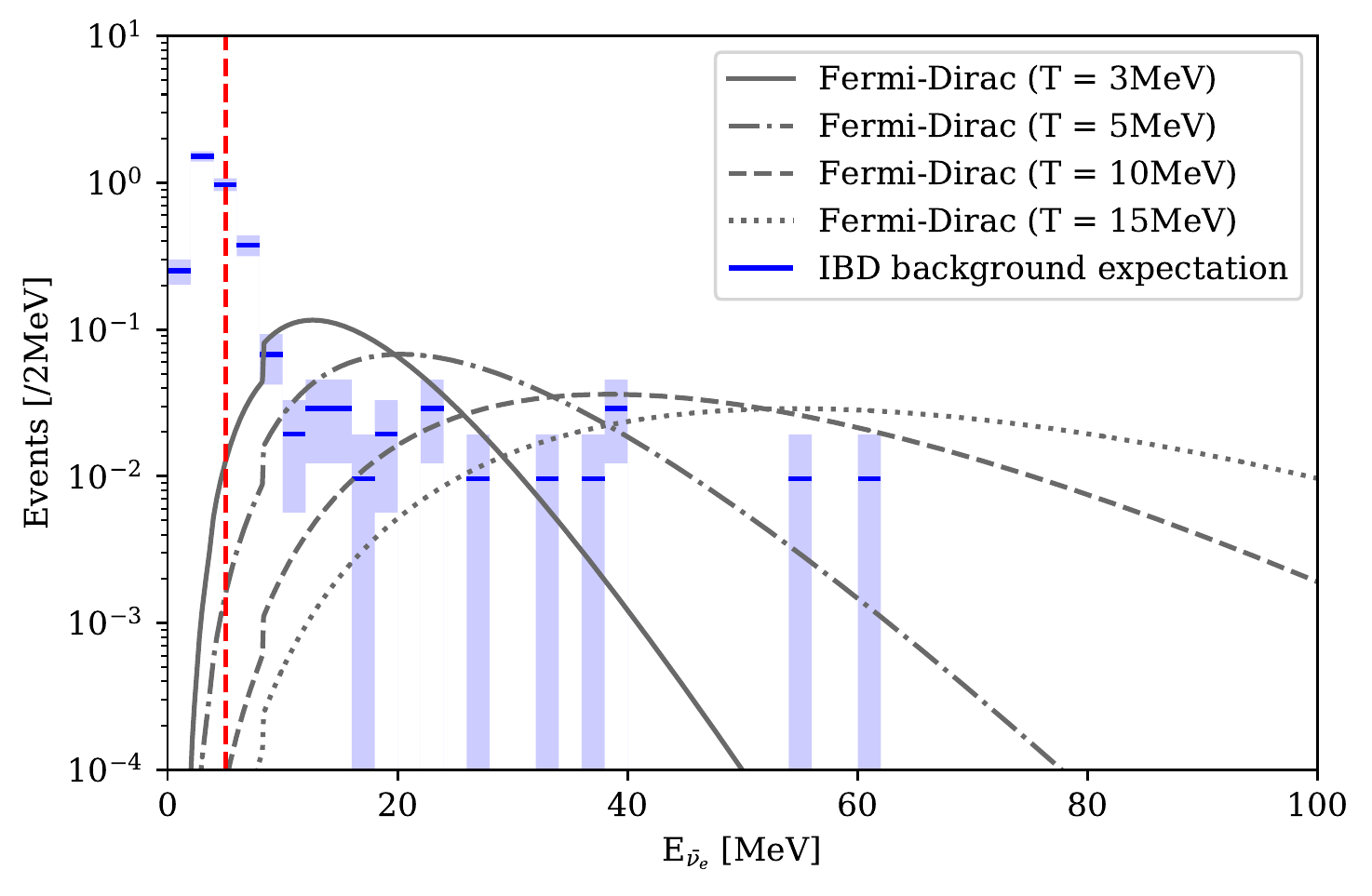}{0.508\textwidth}{}
    }
    \caption{Left: The constraint on the \nue source luminosity and effective temperature relation in terms of LGRB (thin solid blue), SGRB (dotted blue), and the combined set (thick solid blue). The SN1987A neutrino measurements from the Kamiokande-II~\citep{hirata1988observation}, IMB~\citep{schramm1987neutrinos}, and Baksan~\citep{lattimer1989analysis} experiments are shown in solid, dashed, and dotted closed red contours, respectively. 
    Right: The \nue energy distribution background expectation throughout the four KamLAND periods are shown in blue with the 1\,$\sigma$ statistical errors. Assuming a FD neutrino energy distribution from the GRB source, we show the 90\% C.L. upper limit at 3\,MeV, 5\,MeV, 10\,MeV, and 15\,MeV \nue source temperature. The vertical dashed red line corresponds to the energy of the single observed coincident event.
    }
    \label{fig::Lum}
\end{figure}

We now use $N_{90}$ and the subset of GRBs with a measured redshift, $z$, to constrain the \nue integrated luminosity, $L$, and effective \nue temperature, $T$. The \nue spectrum from a single source can be written in terms of a luminosity as: 
\begin{equation}
\psi(E_{\bar{\nu}_e},T,L)_{FD} = \frac{L}{\langle{}E\rangle{}} \lambda(E_{\bar{\nu}_e},T)_{FD} .
\end{equation}
The expected total flux at the detector in period $k$ is therefore:
\begin{equation}
\Psi^k(E_{\bar{\nu}_e},T,L) = \sum_i^{i\in k} \frac{1 + z_i}{4\pi d_i^4} \psi((1+z_i) E_{\bar{\nu}_e},T,L),
\end{equation}
where $z_i$ and $d_i$ are the redshifts and luminosity distances to the $i^{\mathrm{th}}$ GRB. The \nue effective temperature and luminosity upper limits ($T_{up}$, $L_{up}$) are then connected to $N_{90}$ through:
\begin{equation}\label{eq::lum}
N_{90} = N_T  \sum_{k=1}^{k=4} \int_{E^k_{low}}^{E_{high}}  \epsilon_{live}^k \epsilon^k_s(E_{\bar{\nu}_e}) \sigma(E_{\bar{\nu}_e}) \Psi^k(E_{\bar{\nu}_e},T_{up},L_{up}) dE_{\bar{\nu}_e},
\end{equation}
where $E_{high}= 100$\,MeV for $k$\,=\,Period I\,-\,IV, $E^k_{low}= 1.8$\,MeV for $k$ in Period II\,-\,IV, and 8.3\,MeV for Period I. The \nue source luminosity upper limit at a given effective temperature  is plotted in Fig.~\ref{fig::Lum} (Left). This figure also includes the 68\% C.L. compatible regions (red) from the observed neutrinos from SN1987A; these measurements originate from the Kamiokande-II\footnote{It has been suggested that the Kamiokande-II observation of SN1987A is perhaps a factor of two low due to an unrecognized tape drive error and reset problem, which may have caused the much discussed 7.3\,s gap in the Kamiokande-II data. If correct, this would make the overlap with the IMB and Baksan data more consistent. See \cite{oyama2021comment}.}~\citep{hirata1988observation}, IMB~\citep{schramm1987neutrinos}, and Baksan~\citep{lattimer1989analysis} experiments. 

Fig.~\ref{fig::Lum} (right) shows the expected background spectra (blue histogram) from the four KamLAND periods. 
We also include the shape of the FD neutrino spectrum at $T_{up}$\,=\,3\,MeV,\,5\,MeV,\,10\,MeV,\,and\,15\,MeV, with the corresponding luminosity, $L_{up}$ from Eq.~(\ref{eq::lum}).

As noted in Sec.~\ref{sec::intro}, GRB170817A was determined to be class SGRB. 
Since this event is also the closest observed GRB, $\approx$\,40\,Mpc, it also is the dominant contribution to the limit calculated through Eq.~(\ref{eq::lum}). We therefore placed this event into the SGRB class for this analysis. It has been noted, that GRB170817A also lacks the hard spectrum of a SGRB and may indicate additional classes or subclasses in the current GRB classification scheme~\citep{horvath2018classifying}.

The results presented here assume an isotropic neutrino emission from the source, however, the accretion disk and remnant are expected to radiate neutrinos preferentially along the polar direction~\citep{perego2014neutrino}. Even so, the limits indicate that KamLAND should not expect any significant correlated observation until the observation of a GRB localized to the Milky way or one of the Milky Way's satellite galaxies \textbf{within $\sim$0.5\,Mpc}. We also note that the calculations do not incorporate neutrino oscillations or other flavor changing effects that could either increase or decrease the number of neutrinos observed by KamLAND.

If a GRB is localized \textbf{within $\sim$0.5\,Mpc}, a future search should be conducted on that individual source rather than performing a stacked analysis, as was have performed here. If the background expectation in the search time window is below 0.50 (0.12) events, a single coincident observation would yield a non-zero 90\% (99\%) C.L. lower limit on the electron antineutrino flux. If this analysis is repeated with a significantly larger detector, and thus a larger background expectation, the limits can be determined separately for each energy bin, as in~\cite{orii2021search}. Alternatively, a low-energy cutoff at $\sim$7.5\,MeV can be placed, as we have done for Period I, to reduce the reactor neutrino background. The background estimation for these analyses will also benefit from tighter constraints on the predefined window size t$_p$. This will primarily be influenced by revised limits on the sum of the neutrino masses or a measurement of an effective neutrino mass, and potentially theoretical model dependent time differences between the neutrino and photon production. Finally, if the GRB happens to occur nearby, a separate search for the less likely NC \nue interaction with $^{12}$C at 15\,MeV could be performed with very low background levels, expected to be dominated by atmospheric neutrino interactions.



\section{Conclusion} \label{sec:conclusion}
This paper presents a time-coincident search for low-energy $\bar{\nu}_e$s in the KamLAND detector with \textcolor{\cc}{2795} LGRBs and \textcolor{\cc}{465} SGRBs. We search for events between the first GRB 2004 December 19th and the most recent KamLAND run in 2021 June 12th using a GRB catalog compiled from the GCN and GBM. With a time window of $\pm$500\,s around each GRB plus the GRB event duration, we find a single candidate \nue coincident event. This observation is not statistically significant. From this, we present an upper limit on the \nue fluence, placing the most stringent limit below \textcolor{\cc}{17.5\,MeV}. Finally, using the known redshifts to the subset of GRBs, we assume a Fermi-Dirac energy spectrum to place a limit on the \nue luminosity and effective temperature. 

\begin{acknowledgments}

The KamLAND experiment is supported by JSPS KAKENHI Grants 19H05803; the World Premier International Research Center Initiative (WPI Initiative), MEXT, Japan; Netherlands Organization for Scientific Research (NWO); and under the U.S. Department of Energy (DOE) Contract No.~DE-AC02-05CH11231, the National Science Foundation (NSF) No.~NSF-1806440,~NSF-2012964, the Heising-Simons Foundation, as well as other DOE and NSF grants to individual institutions. The Kamioka Mining and Smelting Company has provided services for activities in the mine. We acknowledge the support of NII for SINET4. 

\end{acknowledgments}

\bibliography{Main}{}
\bibliographystyle{aasjournal}

\end{document}